\documentclass[lettersize,journal]{IEEEtran}
\usepackage{amsmath,amsfonts}
\usepackage{algorithmic}
\usepackage{algorithm}
\usepackage{array}
\usepackage[caption=false,font=normalsize,labelfont=sf,textfont=sf]{subfig}
\usepackage{textcomp}
\usepackage{stfloats}
\usepackage{url}
\usepackage{verbatim}
\usepackage{graphicx}
\usepackage{cite}
\usepackage{soul}
\usepackage{xcolor}
\usepackage{kotex}
\usepackage{xcolor}
\usepackage[switch]{lineno} 

\begin{document}

\title{Frequency-Division Multiplexed CV-QKD System}

\author{Jaehyeok Han$^{\dagger}$, 
        Donghyeok Lee$^{\dagger}$, 
        Minseok Ryu,
        Syed Assad,
        Yong-Su Kim$^{*}$, and
        Sunghyun Bae$^{*}$,~\IEEEmembership{Senior Member, IEEE}

\thanks{Manuscript received March 20, 2026. This work was supported in part by IITP grant (No. RS-2024-00396999), NRF grant (No. RS-2023-00242396), and NIA funded by the Ministry of Science and ICT(MSIT, Korea). [Quantum technology test verification and consulting support in 2024].}
\thanks{J. Han and D. Lee are with the Department of Information and Communications Engineering, Sejong University, Seoul 05006, South Korea, and also with the Center for Quantum Technology, Korea Institute of Science and Technology (KIST), Seoul 02792, South Korea (e-mail: hanjaehyeok@sju.ac.kr; ldh0906@sju.ac.kr).}
\thanks{M. Ryu and Y.-S. Kim are with the Center for Quantum Technology, Korea Institute of Science and Technology (KIST), Seoul 02792, South Korea (e-mail: rminseok99@gmail.com; yong-su.kim@kist.re.kr).}%
\thanks{S. Assad is with the Quantum Innovation Center (Q. InC), Agency for Science, Technology and Research (A*STAR), Singapore 138634 (e-mail: cqtsma@gmail.com).}%
\thanks{S. Bae is with the Department of Quantum Information Science and Engineering, Sejong University, Seoul 05006, South Korea  and also with the Department of Information and Communications Engineering, Sejong University, Seoul 05006, South Korea (e-mail: sungbae@sejong.ac.kr).}%
\thanks{$^{\dagger}$J. Han and D. Lee contributed equally to this work.}
\thanks{$^{*}$Corresponding authors: Yong-Su Kim and Sunghyun Bae.}%
}

\markboth{IEEE PHOTONICS TECHNOLOGY LETTERS,~Vol.~XX, No.~XX, XX~2026}%
{Shell \MakeLowercase{\textit{et al.}}: A Sample Article Using IEEEtran.cls for IEEE Journals}


\maketitle

\begin{abstract}
We propose a frequency-division multiplexed (FDM) continuous-variable quantum key distribution (CV-QKD) system with enhanced spectral efficiency through optimized channel spacing of low-symbol-rate signals. A four-channel 10-Mbaud FDM-CV-QKD system was experimentally demonstrated using Gaussian modulation, a transmitted local oscillator, and homodyne detection. Despite the inter-channel interference, under a finite-size scenario ($m=1.25\times10^6$), the system achieved a 3.6-fold back-to-back secret key rate gain and outperformed the single-channel frequency-upconverted signal up to 26.8~km.
\end{abstract}


\section{Introduction}
\IEEEPARstart{Q}{uantum} key distribution (QKD) enables two distant parties to establish a shared secret key whose security is guaranteed by the fundamental laws of quantum physics \cite{84_IEEE_BB84}. The no-cloning theorem ensures that any eavesdropping attempt inevitably disturbs the quantum states, allowing the legitimate parties to detect the presence of an adversary. Since the pioneering BB84 protocol~\cite{84_IEEE_BB84}, various discrete-variable (DV) protocols have been developed and experimentally demonstrated~\cite{92_PRL_Bennett, 04_PRL_Gobby}. However, DV-QKD systems typically require single-photon sources and detectors, which are incompatible with standard optical communication infrastructure.

Continuous-variable (CV) QKD offers a practical alternative by encoding information onto the quadratures of the electromagnetic field \cite{99_PRA_T_Ralph, 02_PRL_F_Grosshans}. The security of CV-QKD is rooted in the Heisenberg uncertainty principle, which prevents an eavesdropper from simultaneously measuring conjugate quadratures without introducing noise \cite{05_arXiv_Koashi}. Early CV-QKD schemes required entangled EPR states, posing significant experimental challenges \cite{99_PRA_T_Ralph}. The GG02 protocol \cite{02_PRL_F_Grosshans} resolved this difficulty by using Gaussian-modulated coherent states \cite{03_Nature_F_Grosshans}, enabling the use of standard telecom components such as a coherent laser and homodyne detectors. This makes CV-QKD a cost-effective and potentially high-key-rate solution, particularly attractive for short- and mid-haul networks.

Multi-carrier CV-QKD has emerged as a promising architecture due to its potential scalability to multiparty key distribution in point-to-multipoint topologies and its resilience against localized spectral disturbances in point-to-point topologies \cite{14_PRA_J_Fang, 18_CSF_L_Gyongyosi, 23_OE_Wang, 25_Optica_H_Wang}. This approach also offers practical advantages such as simplifying pilot-clock integration for phase-noise tracking \cite{25_Optica_H_Wang}. Although orthogonal frequency division multiplexing (OFDM) is an efficient multi-carrier technique for enhancing spectral efficiency, it requires complex digital signal processing based on the fast Fourier transform (FFT) and inverse FFT (IFFT) for all subcarriers, along with stringent synchronization across multiple carriers  \cite{23_OE_Wang, 25_Optica_H_Wang}. In contrast, FDM allows each subcarrier to be processed independently, enabling cost-effective user terminals in point-to-multipoint configurations \cite{14_PRA_J_Fang, 18_CSF_L_Gyongyosi}.

In this Letter, we propose an FDM-based approach for cost-effective multi-channel CV-QKD and present a proof-of-principle demonstration of a $4\times10$~Mbaud FDM-CV-QKD system. While our experimental demonstration employed Gaussian modulation, a transmitted local oscillator (TLO), and homodyne detection, the proposed scheme is equally applicable to local LO (LLO) and heterodyne detection. The system utilizes pilot clocks to implement basis selection. By increasing the number of FDM channels from one to four, we show that the total secret key rate (SKR) scales favorably. Furthermore, we demonstrate that closely-spaced multiplexing of low-symbol-rate signals near the crosstalk-suppression limit is more advantageous for the SKR of the FDM system than employing fewer high-symbol-rate channels, as it concentrates the channels within the frequency range where shot-noise dominance is maintained.

\section{Experiment and Results}
\begin{figure}[!t]
\centering
\includegraphics[width=0.9\columnwidth]{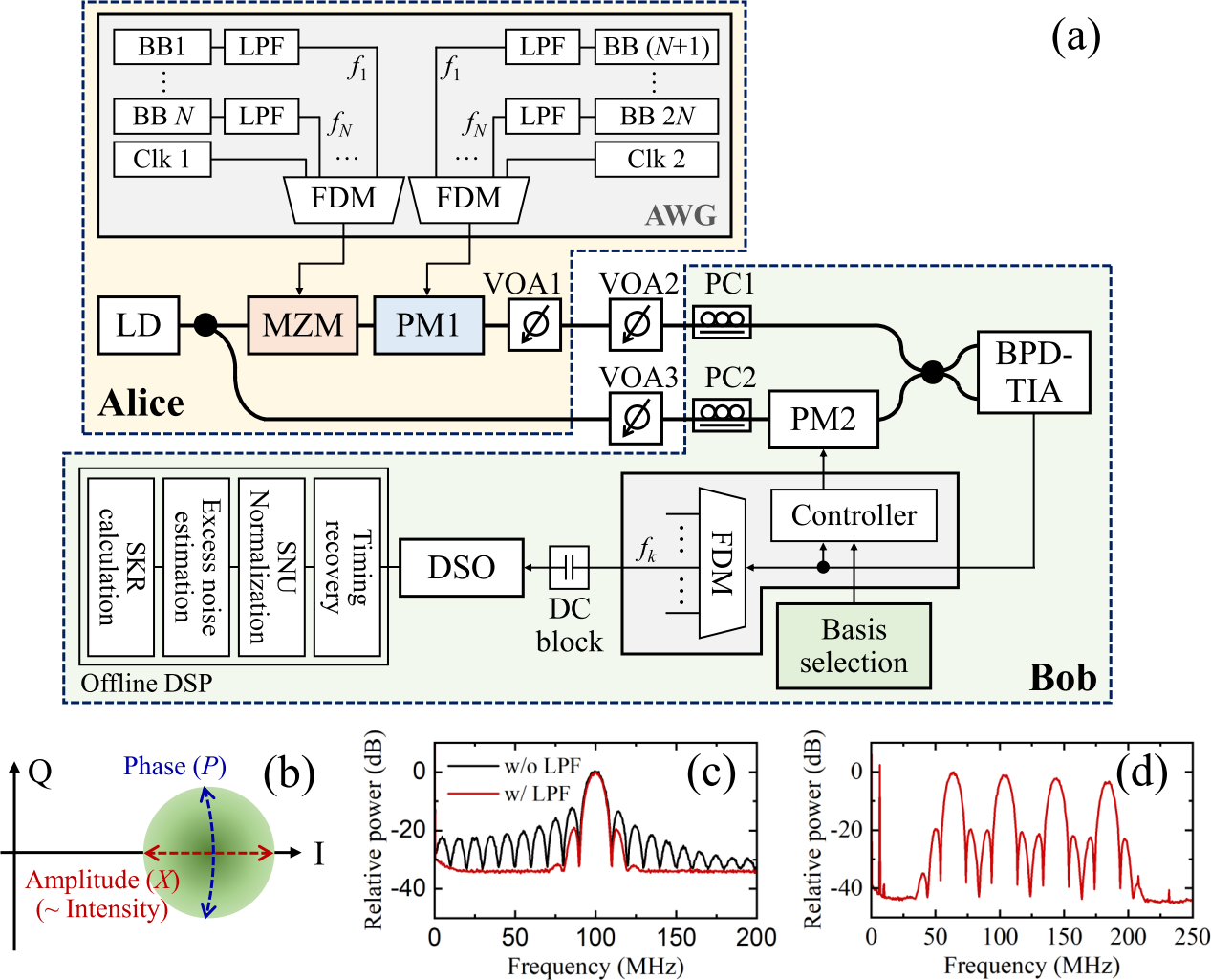}
\caption{(a) Experimental setup for the FDM-CV-QKD system (BB: baseband signal, Clk: clock). (b) Schematic illustration of the Gaussian-modulated CV-QKD signal distribution in phase space. Normalized electrical spectra of (c) an IF signal ($f_{\text{IF}} = 100$~MHz) generated from a 10-Mbaud baseband signal, with and without a 10-MHz LPF applied, and (d) the 4-channel FDM signal (10~Mbaud per channel, initial $f_{\text{IF}}=64$~MHz, channel spacing = 40~MHz) with a 7-MHz pilot clock. }
\label{Fig1}
\end{figure}

Fig.~1(a) shows the experimental setup for the proposed FDM-based CV-QKD system. The output of a 1560-nm laser (linewidth = 600 kHz) was split into two paths. One was used as an LO (with 7-dBm optical power to ensure shot-noise-limited detection), and the other was modulated to generate the CV-QKD signal using an $x$-cut Mach-Zehnder modulator (MZM) with a guaranteed chirp parameter of $-0.1 \leq \alpha \leq 0$ biased at its quadrature point, followed by a phase modulator (PM1). As illustrated in Fig.~1(b), the large coherent displacement relative to the modulation range ensures that the intensity modulation is linear with the field amplitude. Under this condition, the MZM effectively performs amplitude modulation. The driving signals for the MZM and PM1 were generated using an arbitrary waveform generator (AWG; maximum sampling rate = 1.2 GSa/s, bandwidth = 250 MHz) as follows. A total of $2N$ independent 10-Mbaud Gaussian-modulated baseband signals (1 to $N$ for the MZM; $(N+1)$ to $2N$ for PM1) were filtered using fourth-order Bessel low-pass filters (LPF; 3-dB bandwidth = 10 MHz), as shown in Figs.~1(c) and (d), to avoid aliasing from sideband overlap. The filtered signals were then frequency-multiplexed separately for the MZM and PM1, with the $k$-th and $(N+k)$-th signals sharing a common intermediate frequency (IF) of $f_{\text{IF},k}$. All $2N$ baseband signals were scaled to a uniform modulation variance ($\text{V}_\text{mod}$). In this FDM configuration, the sidelobe suppression by the LPFs also mitigated inter-channel crosstalk. For basis selection, 7- and 8-MHz pilot clocks (Clk 1 and 2, with an amplitude of $0.05 \text{V}_\pi$ where $\text{V}_\pi$ is the half-wave voltage) were added to the MZM and PM1 driving signals, respectively. Subsequently, $\text{V}_\text{mod}$ for all subcarriers was calibrated to 5.8 shot noise units (SNU) using a variable optical attenuator (VOA1), after pre-equalizing the AWG amplitude of each subcarrier to ensure a uniform $\text{V}_\text{mod}$ across the band. This value was chosen to maximize the asymptotic secret key rate (SKR) for each subcarrier at a target distance of 20~km, assuming a reverse reconciliation efficiency $\beta$ of 0.9 and an excess noise variance of 0.05 SNU under the trusted-device scenario. The SKR of the $k$-th baseband signal, $R^{(k)}$, under finite-size analysis is given by \cite{10_PRA_A_Leverrier}
\begin{equation}
\begin{aligned}
R^{(k)} = f_{\text{sym}} \frac{n}{N} \left(\beta I_{AB}^{(k)}(m) - \chi_{BE}^{(k)}(m;\bar\epsilon) - \Delta(n;\bar\epsilon)\right),
\end{aligned}
\label{eq1}
\end{equation} \noindent
where $f_{\text{sym}}$ is the symbol rate (10~Mbaud in this experiment), $N$ is the total number of transmitted symbols, consisting of $m$ symbols for parameter estimation and $n$ symbols for raw key generation ($N=m+n$). $I_{AB}^{(k)}(m)$ represents the mutual information between Alice and Bob for the $k$-th channel ($k=1,2,...,2N$) estimated using $m$ symbols, which is determined by the modulation variance $\text{V}_{\text{mod}}^{(k)}$ and the total noise $\Xi_{\text{tot}}^{(k)}$. Under the TLO configuration, $\Xi_{\text{tot}}^{(k)}$ is expressed as
\begin{equation}
\Xi_{\text{tot}}^{(k)} = \frac{1-T_{\text{ch}}}{T_{\text{ch}}} + \epsilon^{(k)} + \frac{1}{T_{\text{ch}}} \left( \frac{1-\eta_{\text{det}}}{\eta_{\text{det}}} + \frac{\nu_{\text{det}}^{(k)}/T_{\text{ch}}}{\eta_{\text{det}}} \right),
\label{eq2}
\end{equation} \noindent 
where $T_{\text{ch}}$ is the channel transmittance, $\epsilon^{(k)}$ is the excess noise variance, and $\eta_{\text{det}}$ is the quantum efficiency of the detector. In the TLO configuration, the detector noise variance $\nu_{\text{det}}^{(k)}$ is effectively scaled by $1/T_{\text{ch}}$ due to shot noise reduction from LO attenuation. $\chi_{BE}^{(k)}(m;\bar\epsilon)$ is the Holevo bound estimated using $m$ symbols and $\Delta(n;\bar\epsilon)$ is the finite-size security correction for privacy amplification, both evaluated at a failure probability of $\bar\epsilon$. As $N$ approaches infinity, $R^{(k)}$ converges to the asymptotic rate:
\begin{equation}
\begin{aligned}
R_{\text{asympt}}^{(k)} = f_{\text{sym}} (\beta I_{AB}^{(k)} - \chi_{BE}^{(k)}). \\
\end{aligned}
\label{eq3}
\end{equation}
Alice sent the generated CV-QKD signal and the LO to Bob through two separate fibers, with the channel loss emulated using VOA2 and VOA3. This spatial separation eliminates the polarization-leakage-induced impairments that may arise when the signal and LO are polarization-multiplexed within a TLO-based system, allowing the FDM-CV-QKD performance to be analyzed independently of such effects. In the experiment, VOA1 and VOA2 were combined into a single VOA.

On Bob's side, the polarizations of the received signal and LO were aligned using polarization controllers (PC1 and PC2). PM2 controlled the phase of the received LO for measurement basis selection (i.e., amplitude or phase). The received QKD signal was detected with the received LO using a homodyne detector consisting of a $2\times2$ optical coupler and a balanced photodetector with an integrated transimpedance amplifier (BPD-TIA) (bandwidth = 500 MHz, $\eta_{\text{det}}$ = 0.83). The amplitude and phase bases were selected by adjusting the PM2 bias to nullify the 8- and 7-MHz pilot clocks, respectively. To detect the \mbox{$k$-th} or \mbox{$(N+k)$-th} signal depending on the basis selection, the received signal was down-converted to baseband using an FDM demultiplexer (bandwidth = 35.16 MHz), and the resulting waveforms were captured by a digital storage oscilloscope (DSO; maximum sampling rate = 6.25~GSa/s). The total secret key rate $R_{T}$ is as follows:  
\begin{equation}
R_{T} = \sum\nolimits_{k=p}^{p+N-1} R^{(k)} ,
\label{eq4}
\end{equation}
where $p$ is set to 1 or $N+1$ for amplitude or phase basis, respectively.

\begin{figure}[!t]
\centering
\includegraphics[width=\columnwidth]{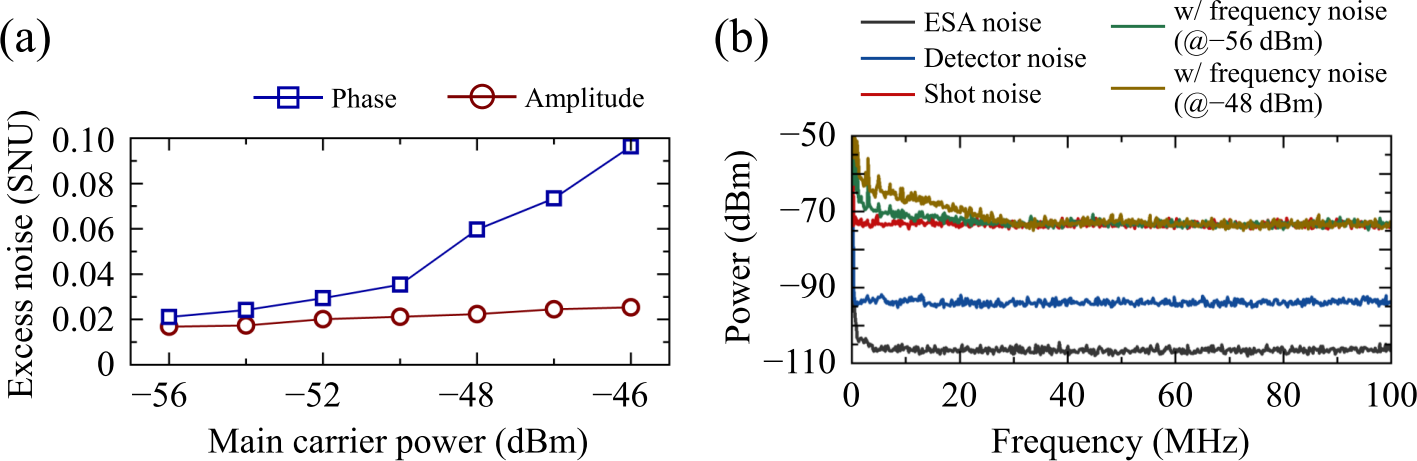}
\caption{(a) Estimated excess noise as a function of the main carrier power of the CV-QKD signal. (b) Measured spectra of detector noise, shot noise, and total noise including frequency noise at main carrier powers of -56 and -48~dBm. }
\label{Fig2}
\end{figure}

The proposed system generated CV-QKD signals modulated on subcarriers at $f_\text{IF}$, which are displaced by the main carrier. As illustrated in Fig.~1(b), when the main carrier power is sufficiently large and the modulation depth is small, the amplitude and phase modulations can be approximated as in-phase and quadrature (I/Q) modulations, respectively. This approximation may not hold if the main carrier power is not sufficiently large relative to the modulation depth, which would be reflected in the excess noise. 

To determine the optimal operating conditions, we evaluated the excess noise of a single-channel 10-Mbaud IF signal ($f_\text{IF}=100$~MHz) as a function of the main carrier power using $1.25\times10^6$ symbols, as shown in Fig.~2(a). As the main carrier power increased, both the amplitude- and phase- basis excess noise increased due to frequency noise induced by the main carrier, the spectra of which (particularly in the 0--20 MHz range) are shown in Fig.~2(b). It is worth noting that the phase-basis excess noise increased much more sharply, since the phase-modulated component is more directly susceptible to frequency noise; this susceptibility was further amplified by the reduced modulation depth at higher power levels. Consequently, the main carrier power was set to $-56$~dBm to sufficiently reduce the total excess noise. This power level corresponds to a mean photon number of 2000, which is sufficient to ensure the small-modulation approximation for generating the QKD signal on the subcarrier ($\text{V}_{\text{mod}}=5.8$~SNU, corresponding to a photon number variance of 2.9). 

\begin{figure}[!t]
\centering
\includegraphics[width=\columnwidth]{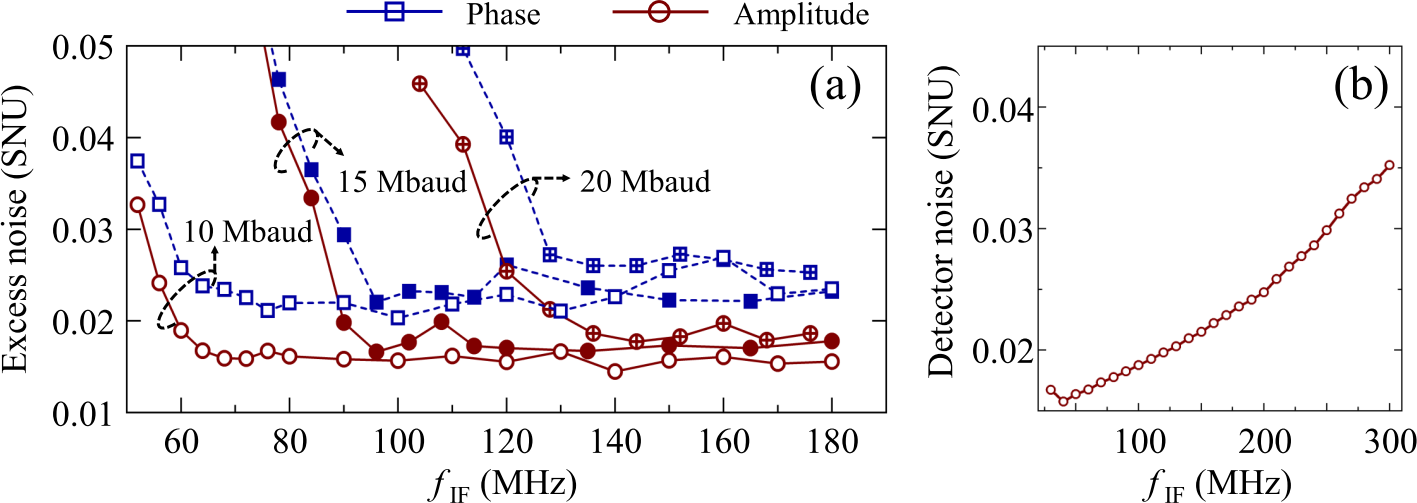}
\caption{(a) Estimated excess noise as a function of $f_{\text{IF}}$ for 10-, 15-, and 20-Mbaud single-channel signals. (b) Ratio of detector noise to shot noise at various $f_{\text{IF}}$, measured with a 10-MHz noise bandwidth. }
\label{Fig3}
\end{figure}

We investigated the relationship between the baseband symbol rate and $f_\text{IF}$ required to suppress excess noise caused by sideband overlap. Fig.~3(a) shows the estimated excess noise as a function of $f_\text{IF}$ for single-channel signals at 10, 15, and 20 Mbaud. For 10-, 15-, and 20-Mbaud signals, minimum $f_\text{IF}$ values of 64, 96, and 136 MHz were necessary to suppress excess noise, establishing the following empirical relationship:
\begin{equation} 
f_{\text{IF}} / {\text{SR}_{\text{BB}}} \geq 6.4 ,
\label{eq5} 
\end{equation}
where $\text{SR}_{\text{BB}}$ is the baseband symbol rate. This ratio is attributed to the spectral components up to the second sidelobe in the filtered baseband signal, as shown in Fig.~1(c). When scaling the system to FDM, this result provides the design criterion for the first channel. It is worth mentioning that the saturated excess noise increased with the symbol rate, as higher symbol rates necessitate higher first-channel frequencies where shot-noise dominance was reduced, as shown in Fig.~3(b).

\begin{figure}[!t]
\centering
\includegraphics[width=\columnwidth]{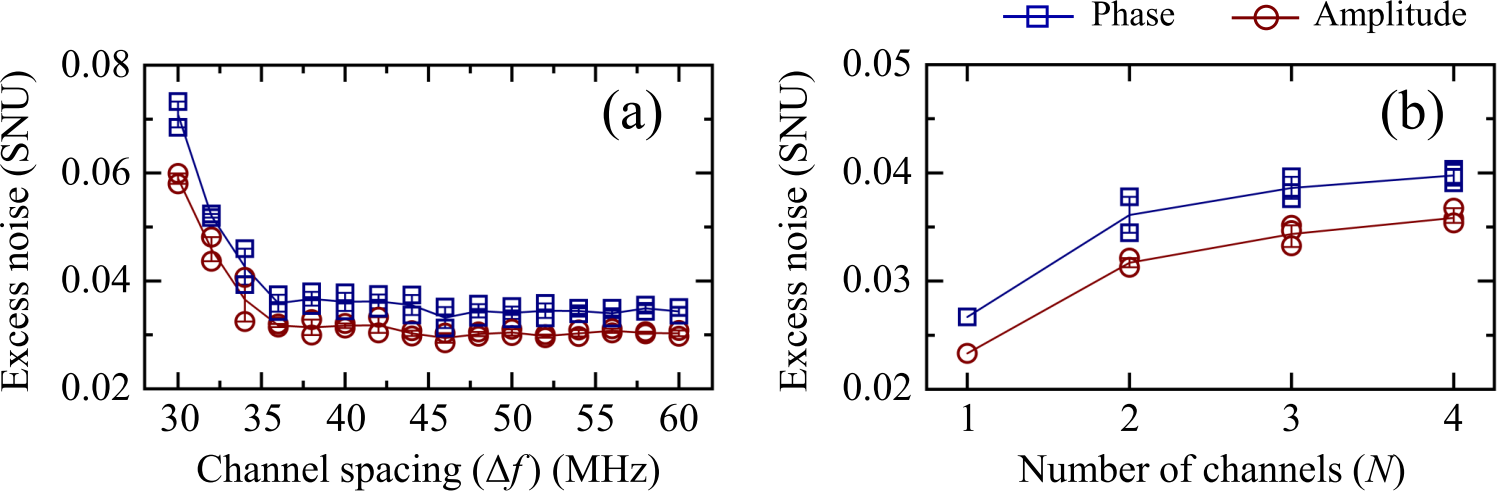}
\caption{(a) Estimated excess noise of each channel versus channel spacing in a two-channel FDM system (10 Mbaud per channel), with the first channel at $f_{\text{IF}}=64$~MHz. (b) Estimated excess noise for all channels as a function of the number of channels, with the first channel at $f_{\text{IF}}=64$~MHz and a 40-MHz channel spacing. Markers represent individual channel excess noise, while lines indicate the average. }
\label{Fig4}
\end{figure}

We then evaluated the impact of FDM on QKD signal performance using 10-Mbaud signals per channel. Although a 64-MHz channel spacing, as implied by Eq.~\eqref{eq5}, would suppress inter-channel crosstalk, we adopted a narrower spacing based on a relaxed criterion to improve spectral efficiency. A two-channel FDM system was investigated with the first channel centered at 64 MHz (based on Eq.~\eqref{eq5}), as shown in Fig.~4(a). For both amplitude- and phase-modulated signals, crosstalk was sufficiently attenuated at a channel spacing ($\Delta f$) of 36~MHz or higher. Consequently, we adopted a channel spacing of 40~MHz, corresponding to $4\times\text{SR}_\text{BB}$, preventing the first-sidelobe overlap. 

Fig.~4(b) shows the excess noise for both amplitude- and phase-modulated signals as the number of channels increased, with the first channel at 64~MHz and a 40-MHz spacing. For amplitude-modulated signals, the excess noise increased by a factor of 1.36 when the number of channels doubled from one to two. However, the rate of noise growth diminished as the number of channels further increased. This was because the spectral distance between existing and newly added channels increased, leading to a reduction in additional crosstalk. Specifically, the excess noise increased by only a factor of 1.13 as the number of channels expanded from 2 to 4. The same trend was observed for phase-modulated signals.

\begin{figure}[!t]
\centering
\includegraphics[width = 0.8\columnwidth]{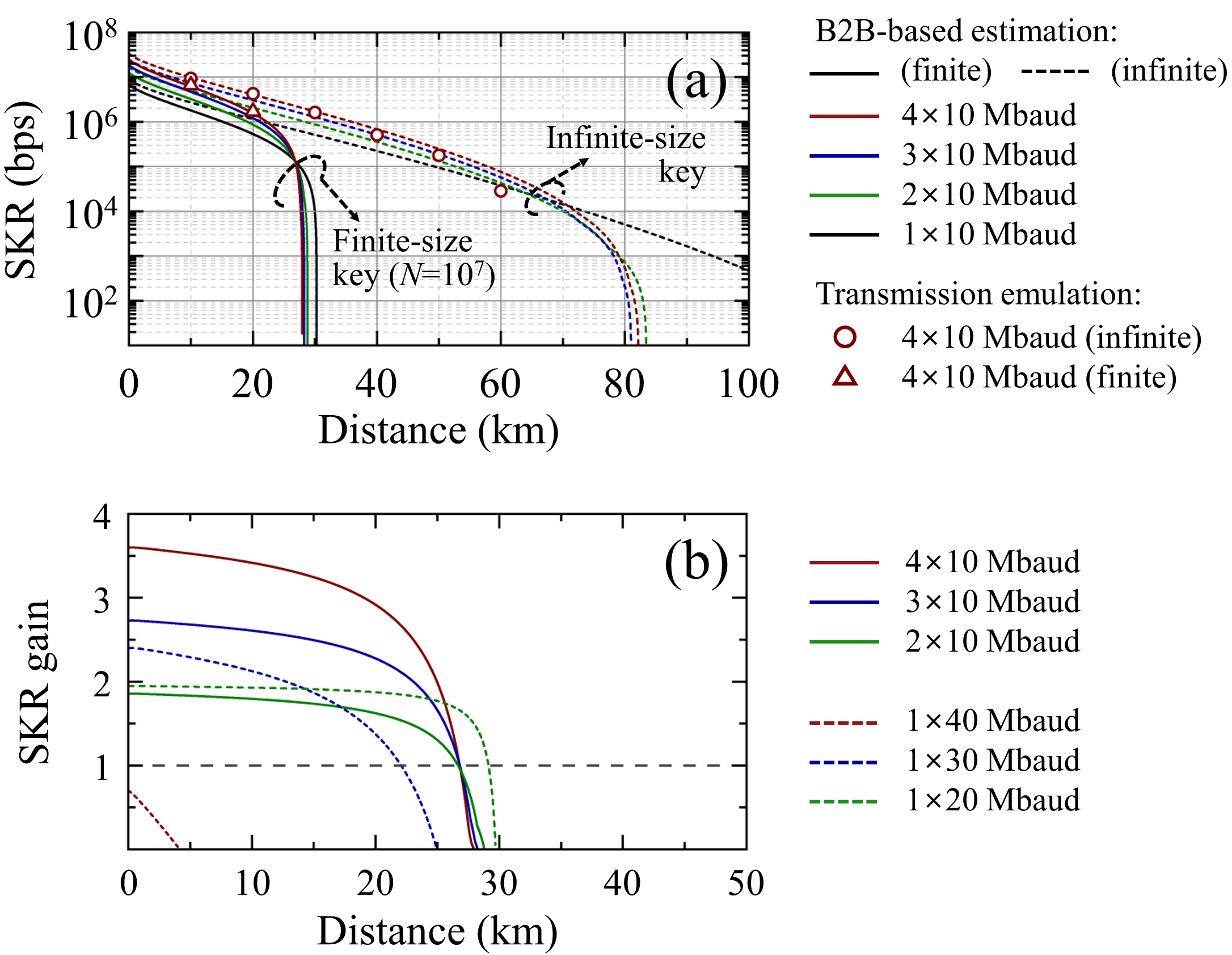}
\caption{(a) Total SKR as a function of transmission distance for FDM-CV-QKD with 10-Mbaud signals, comparing finite- ($N=10^7, m=1.25 \times 10^6$, solid lines) and infinite-size (dashed lines) key scenarios. Markers represent experimental verification via emulated channel loss using VOAs. (b) SKR gain as a function of transmission distance, comparing multi-channel 10-Mbaud FDM signals (solid lines) with single-channel high-baud-rate signals (dashed lines). The line-style conventions in (a) and (b) are defined independently within each panel.}
\label{Fig5}
\end{figure}

We experimentally investigated the achievable SKR, adopting a worst-case approach by selecting the lower value between the amplitude and phase bases, as a function of transmission distance based on the back-to-back excess noise (estimated with $m=1.25\times10^6$ symbols), as shown in Fig.~5(a). The analysis considered 1 to 4 channels of 10-Mbaud signals (initial $f_{\text{IF}}=64$~MHz, 40-MHz spacing) with an assumed fiber loss of 0.2~dB/km. In the asymptotic limit (i.e., infinite key size), a single channel could reach up to 119.0~km, whereas the maximum distance for 2--4 channel systems decreased significantly to 81.1--83.5~km. This degradation originates from the increased excess noise with the number of channels, as analyzed in Fig.~4(b). Under this asymptotic scenario, the 4-channel system outperformed the single-channel system for distances up to 71.0~km. However, under a practical finite-size scenario ($(m,n)=(1.25\times10^6 ,8.75\times10^6)$, $\bar\epsilon=10^{-10}$), the transmission distance decreased only slightly from 30.2~km (single-channel) to 27.9~km (four-channel), a penalty of merely 2.3~km. {The corresponding back-to-back SKR gains, summarized in Table}~\ref{tab1}{, scaled nearly proportionally with the number of channels, with a slight sublinear trend arising from the increased excess noise.} Nevertheless, the 4-channel configuration maintained a higher SKR for distances up to 26.8~km, which is comparable to the maximum transmission distance of the single-channel system. 

{As summarized in Table}~\ref{tab1}{, the up-converted single-channel signals, with the 20-, 30-, and 40-Mbaud signals frequency-shifted to $f_{\text{IF}}=128$, 192, and 256~MHz, respectively, performed comparably to the FDM system only at the lowest aggregate symbol rate and fell increasingly behind as the symbol rate increased.} Notably, a single 30-Mbaud signal exhibited a much lower SKR than a 3-channel FDM system occupying the same passband bandwidth. This is because the 30-Mbaud signal requires a higher $f_\text{IF}$, forcing the system to operate in a high-frequency region where the BPD's shot-noise dominance is significantly diminished. Furthermore, the 40-Mbaud signal consistently yielded a lower SKR than the 10-Mbaud signal, likely due to the bandwidth limitations of the AWG and demultiplexer, alongside the reduced shot-noise-to-electronic-noise ratio.

\begin{table}[!t]
\caption{{Back-to-Back SKR Gain Under Finite-Size ($N=10^{7}$, $m=1.25\times10^{6}$, $\bar\epsilon=10^{-10}$) and Asymptotic Scenarios}}
\label{tab1}
\centering
\begin{tabular}{c cc cc}
\hline\hline
Aggregate & \multicolumn{2}{c}{Gain of FDM (10~Mbaud/ch)} & \multicolumn{2}{c}{Gain of single channel} \\
\cline{2-3}\cline{4-5}
symbol rate & Finite & Asymptotic & Finite & Asymptotic \\
\hline
20 Mbaud & 1.9 & 1.9  & 1.9 & 1.9 \\
30 Mbaud & 2.7 & 2.8  & 2.4 & 2.4 \\
40 Mbaud & 3.6 & 3.7 & 0.7 & 0.8 \\
\hline\hline
\end{tabular}
\\[3pt]
{\raggedright \footnotesize
All gains are normalized to the $1\times$10-Mbaud single-channel system. \par}
\end{table}

\section{Summary}
We proposed an FDM-based approach that enables cost-effective user terminals in point-to-multipoint architectures through independent subcarrier processing. Although such an FDM system can be vulnerable to inter-channel crosstalk, we experimentally demonstrated that the excess noise could be reduced to a sufficiently low level---enabling secret key generation---with a channel spacing of four times the symbol rate of each subcarrier. Through a 1- to 4-channel demonstration, we further confirmed that the additional crosstalk-induced excess noise diminished as more channels were added, owing to the increased spectral distance between newly added and existing channels; this suggests that the FDM architecture can be extended to more channels, within the available device bandwidth. We also showed that increasing the number of channels while reducing the per-subcarrier symbol rate further enhances the spectral efficiency of the FDM system: the $4\times10$-Mbaud signal achieved a higher SKR than the $1\times40$-Mbaud passband signal across all transmission distances, exhibited a 3.6-fold back-to-back SKR gain over the single-channel system, and reached a maximum transmission distance of 26.8~km. The present analysis applies equally to an LLO configuration, which offers stronger security{, since replacing the TLO with an LLO changes the shot-noise dominance only in a spectrally flat manner across all frequencies; the LLO also avoids the LO loss at longer distances, slightly improving the shot-noise dominance and thus potentially even increasing the SKR, provided that the phase noise is adequately compensated.} A real-fiber demonstration with LLO and appropriate phase recovery remains an important direction for future work, as additional noise sources such as Raman and Rayleigh scattering would increase the overall excess noise.

\bibliographystyle{IEEEtran}
\bibliography{sample}

@article{84_IEEE_BB84,
  title={Quantum cryptography: Public key distribution and coin tossing},
  author={Bennett, Charles H and Brassard, Gilles},
  journal={Proceedings of IEEE International Conference on Computers, Systems and Signal Processing},
  pages={175--179},
  year={1984},
  address={Bangalore, India},
  publisher={IEEE}
}

@article{92_PRL_Bennett,
  title={Experimental quantum cryptography},
  author={Bennett, Charles H. and Bessette, Fran{\c{c}}ois and Brassard, Gilles and Salvail, Louis and Smolin, John},
  journal={Physical Review Letters},
  volume={68},
  number={21},
  pages={3121--3124},
  year={1992},
  publisher={APS}
}

@article{99_PRA_T_Ralph,
  title={Continuous variable quantum cryptography},
  author={Ralph, Timothy C},
  journal={Physical Review A},
  volume={61},
  number={1},
  pages={010303},
  year={1999},
  publisher={APS}
}

@article{02_PRL_F_Grosshans,
  title={Continuous variable quantum cryptography using coherent states},
  author={Grosshans, Fr{\'e}d{\'e}ric and Grangier, Philippe},
  journal={Physical Review Letters},
  volume={88},
  number={5},
  pages={057902},
  year={2002},
  publisher={APS}
}

@article{03_Nature_F_Grosshans,
  title={Quantum key distribution using {G}aussian-modulated coherent states},
  author={Grosshans, Fr{\'e}d{\'e}ric and Van Assche, Gilles and Wenger, J{\'e}r{\^o}me and Brouri, Rosa and Cerf, Nicolas J and Grangier, Philippe},
  journal={Nature},
  volume={421},
  number={6920},
  pages={238--241},
  year={2003},
  publisher={Nature Publishing Group UK London}
}

@article{04_PRL_Gobby,
  title={Quantum key distribution over 122 km of standard telecom fiber},
  author={Gobby, C. and Yuan, Z. L. and Shields, A. J.},
  journal={Physical Review Letters},
  volume={92},
  number={5},
  pages={050503},
  year={2004},
  publisher={APS}
}

@article{05_arXiv_Koashi,
  title={Simple security proof of quantum key distribution via uncertainty principle},
  author={Koashi, Masato},
  journal={arXiv preprint arXiv:quant-ph/0505108},
  year={2005}
}

@article{10_PRA_A_Leverrier,
  title={Finite-size analysis of a continuous-variable quantum key distribution},
  author={Leverrier, Anthony and Grosshans, Fr{\'e}d{\'e}ric and Grangier, Philippe},
  journal={Physical Review A—Atomic, Molecular, and Optical Physics},
  volume={81},
  number={6},
  pages={062343},
  year={2010},
  publisher={APS}
}

@article{14_PRA_J_Fang,
  title={Multichannel parallel continuous-variable quantum key distribution with Gaussian modulation},
  author={Fang, Jian and Huang, Peng and Zeng, Guihua},
  journal={Physical Review A},
  volume={89},
  number={2},
  pages={022315},
  year={2014},
  publisher={American Physical Society},
  doi={10.1103/PhysRevA.89.022315}
}

@article{23_OE_Wang,
  title={Performance analysis for {OFDM}-based multi-carrier continuous-variable quantum key distribution with arbitrary modulation protocol},
  author={Wang, Heng and Pan, Yan and Shao, Yun and Pi, Yaodi and Ye, Ting and Li, Yang and Zhang, Tao and Liu, Jinlu and Yang, Jie and Ma, Li and Huang, Wei and Xu, Bingjie},
  journal={Optics Express},
  volume={31},
  number={4},
  pages={5577--5591},
  year={2023},
  publisher={Optical Society of America}
}

@article{25_Optica_H_Wang,
  title={High-rate continuous-variable quantum key distribution over 100 km fiber with composable security},
  author={Wang, Heng and Li, Yang and Ye, Ting and Ma, Li and Pan, Yan and Wu, Mingze and Li, Junhui and Bian, Yiming and Shao, Yun and Pi, Yaodi and Yang, Jie and Liu, Jinlu and Sun, Ao and Huang, Wei and Pirandola, Stefano and Zhang, Yichen and Xu, Bingjie},
  journal={Optica},
  volume={12},
  number={10},
  pages={1657--1667},
  year={2025},
  publisher={Optica Publishing Group},
  doi={10.1364/OPTICA.566359}
}

@article{18_CSF_L_Gyongyosi,
  title={Multiple access multicarrier continuous-variable quantum key distribution},
  author={Gyongyosi, Laszlo and Imre, Sandor},
  journal={Chaos, Solitons \& Fractals},
  volume={114},
  pages={491--505},
  year={2018},
  publisher={Elsevier}
}

\end{document}